\documentclass[aps,prl,twocolumn,groupedaddress,showpacs]{revtex4} 
\usepackage{graphicx}
\usepackage{dcolumn}
\usepackage{bm}

\begin{document}

\title{Quantum oscillations in a topological insulator Bi$_{1-x}$Sb$_{x}$}

\author{A. A. Taskin}
\author{Yoichi Ando}
\affiliation{Institute of Scientific and Industrial Research, 
Osaka University, Osaka 567-0047, Japan}


\begin{abstract}

We have studied transport and magnetic properties of Bi$_{1-x}$Sb$_x$,
which is believed to be a topological insulator --- a new state of
matter where an insulating bulk supports an intrinsically metallic
surface. In nominally insulating Bi$_{0.91}$Sb$_{0.09}$ crystals, we
observed strong quantum oscillations of the magnetization and the
resistivity originating from a Fermi surface which has a clear
two-dimensional character. In addition, a three-dimensional Fermi
surface is found to coexist, which is possibly due to an unusual
coupling of the bulk to the surface. This finding demonstrates that
quantum oscillations can be a powerful tool to directly probe the novel
electronic states in topological insulators.

\end{abstract}




\maketitle

Bi and Bi$_{1-x}$Sb$_x$ alloys have been ``gold mines" of modern
solid-state physics for over seven decades
\cite{Fuk,Wol,Len,VSE,Br,Ben,Ong}. The discovery of quantum oscillations
itself has been made on pure Bi as long ago as 1930. The anomalously
large diamagnetism in Bi was a fundamentally important problem
\cite{Fuk} and prompted the application of the Dirac theory to solid
states \cite{Fuk,Wol} in the 1960's. Also, Bi$_{1-x}$Sb$_x$ is known as
one of the best thermoelectric materials \cite{Len} and consequently has
been studied in great detail. As a result, Bi and Bi$_{1-x}$Sb$_x$ have
been thought to be very well understood. Yet, a new twist came recently
\cite{K1,K2,MB,SCZ1,K3,SCZ2,MurN,Mol,K4,SCZ3,H1,H2,Nish} from new ideas
about topological phases in condensed matter
\cite{K1,K2,MB,SCZ1,K3,SCZ2,MurN}; namely, it has been realized that
Bi$_{1-x}$Sb$_x$ in the insulating regime (0.07 $< x <$ 0.22) is a
``topological insulator" (TI) that belongs to a non-trivial $Z_2$
topological class \cite{K2}. Since the vacuum belongs to a trivial
topological class, a smooth transition from a TI to the vacuum is
possible only by closing the energy gap along the way, making the
surface of a TI to be intrinsically conducting. Intriguingly, those
metallic surface states are expected to be spin-filtered, obey
Dirac-like energy dispersion, and be topologically protected
\cite{K1,K2,MB,SCZ1}.
 
In the past, a metallic behavior has always been observed at low
temperature in Bi$_{1-x}$Sb$_x$; namely, its resistivity has never
presented a true divergence in the zero-temperature limit, even in the
insulating doping range where a bulk energy gap opens \cite{Len}. The
origin of this metallic state at low temperature has never been fully
understood and was mainly attributed to imperfections of measured
crystals, in particular, to the formation of an impurity band within the
bulk energy gap. On the other hand, recent angle-resolved photoemission
spectroscopy (ARPES) studies on its cleaved (111) trigonal surface have
revealed that the energy dispersions of the surface states possess the
distinctive character to qualify this material as a topological
insulator \cite{K4,SCZ3,H1,H2,Nish}, so it is natural to ask whether the
low-temperature metallic behavior of ``insulating" Bi$_{1-x}$Sb$_x$ has
anything to do with the topological surface state. In the present work,
to directly address this question, we measure quantum oscillations (QOs)
in nominally insulating Bi$_{0.91}$Sb$_{0.09}$ crystal to differentiate
a coherent electronic transport on a fully developed Fermi surface (FS)
from an incoherent transport within an impurity band.

Bi$_{1-x}$Sb$_x$ crystals were grown from a stoichiometric mixture of
99.9999\% purity Bi and Sb elements by a zone melting method. The
homogenization of Bi$_{1-x}$Sb$_x$ microstructure was achieved by
multiple ($\sim$100 times) re-melting of the boule. The last run for
each growth was done at a very low rate ($\le$ 0.2 mm/h) to avoid
constitutional supercooling and the resulting segregation of the solid
solution. All grown crystals were easily cleaved along the (111) plane,
revealing flat, shiny surfaces. To prepare samples suitable for
transport and magnetic measurements, crystals were aligned using the
X-ray Laue analysis and cut along the principal axes. The actual
composition of grown crystals and their purity were checked by the
inductively-coupled-plasma atomic-emission spectroscopy (ICP-AES)
analysis, which confirmed that the composition of grown Bi$_{1-x}$Sb$_x$
was close to the nominal value and that the concentration of any
impurity atoms was less than the sensitivity threshold ($\sim$ 10$^{21}$
m$^{-3}$). The resistivity was measured by a standard four-probe method
on a rectangular sample with the size of approximately
1.5$\times$0.2$\times$0.1 mm$^3$. The electric current was directed
along the $C_3$ axis, and the magnetic field was applied along the $C_2$
axis. For the de Haas-van Alphen (dHvA) measurements, a sample of
approximately cubic shape and the mass of $\sim$80 mg was cut along the
principal crystallographic axes from the same Bi$_{0.91}$Sb$_{0.09}$
crystal as was used for the resistivity measurements. The dc
magnetization $M$ was measured in magnetic fields up to 1 T at fixed
temperatures in the range from 2 to 40 K using a commercial Quantum
Design SQUID magnetometer. Quasi-equilibrium measurement conditions
allowed us to observe even a weak modulation of the magnetization in
fields as low as 0.08 T. 

The temperature dependences of the resistivity, $\rho_{xx}(T)$, for
three high-quality Bi$_{1-x}$Sb$_x$ crystals with increasing Sb
concentration, $x$ = 0.00, 0.06, and 0.09, are shown in Fig. 1(a). A
qualitative change from a metallic behavior with positive slope
$d\rho/dT >$ 0 to an insulator-like behavior is clearly seen between $x$
= 0.06 and 0.09. This indicates that, at $x$ = 0.09, an energy gap for
charge carriers fully opens, in agreement with recent ARPES measurements
\cite{H1,H2,Nish} as well as the well-established evolution of the
energy bands in Bi$_{1-x}$Sb$_x$ upon doping \cite{Len, Br}. The change
from the semimetal nature in pure Bi ($x$ = 0.00), resulting from an
overlap of the valence and conduction bands at different points in the
Brillouin zone (BZ), to the bulk-insulator nature at $x$ = 0.09 with a
direct gap at the zone boundary is schematically depicted in the inset
of Fig. 1(a). Despite the upturn in the resistivity below 200 K, the
sample with $x$ = 0.09 shows a metallic behavior below 40 K, even though
the resistivity value is about 100 times larger than in pure Bi in the
zero-temperature limit.

\begin{figure}\includegraphics*[width=20.5pc]{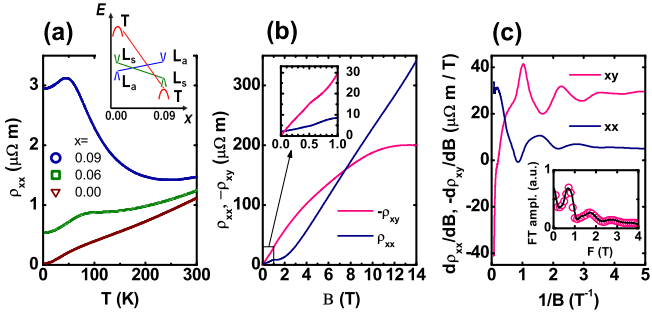}
\caption{(Color online) 
(a) Temperature dependences of $\rho_{xx}$ for $x$ = 0.00, 0.06, 
and 0.09. 
Inset shows a diagram of the evolution of the energy bands 
in Bi$_{1-x}$Sb$_x$ upon doping. 
(b) Magnetic-field dependences of $\rho_{xx}$ and $-\rho_{xy}$ 
in Bi$_{0.91}$Sb$_{0.09}$ at 1.4 K. 
(c) Derivatives of $\rho_{xx}(B)$ and $-\rho_{xy}(B)$ vs. $B^{-1}$.
Inset shows the FT spectrum of the $-\rho_{xy}$ oscillations.
}
\label{fig1}
\end{figure}

Figure 1(b) shows that both $\rho_{xx}$ and $\rho_{xy}$ of the $x$ =
0.09 sample change strongly with the magnetic field $B$ at 1.4 K. The
$\rho_{xx}(B)$ shows an almost linear increase above $\sim$2 T and, at
14 T, it is $\sim$170 times larger than at 0 T. The $\rho_{xy}(B)$, on
the other hand, is almost linear at low fields and shows a saturation
around 14 T. The negative slope of $\rho_{xy}(B)$ at $B$ = 0 T suggests
that the main carriers in the sample are electrons [note that
$-\rho_{xy}(B)$ is plotted in Fig. 1(b)]. The Hall coefficient $R_H$
estimated from the slope is -3.5$\times$10$^{-5}$ m$^{3}$/C, implying
that the concentration of electrons in the sample is of the order of
$|1/eR_H| \approx $ 1.8$\times$10$^{23}$ m$^{-3}$. At low fields, both
$\rho_{xx}(B)$ and $\rho_{xy}(B)$ show a weak modulation in the magnetic
field (more clearly seen in the inset). Their derivatives with respect
to $B$ plotted vs. $1/B$ [Fig. 1(c)] reveal clear Shubnikov-de Haas
(SdH) oscillations in magnetic fields below $\sim$2 T. The Fourier
transform (FT) of $d\rho_{xy}/dB$ [Fig. 1(c) inset] yields the power
spectrum with the leading frequency of 0.67 T. Such an observation of
QOs in a nominally insulating material is surprising, and we elucidate
their origin in the following.

\begin{figure}\includegraphics*[width=20.5pc]{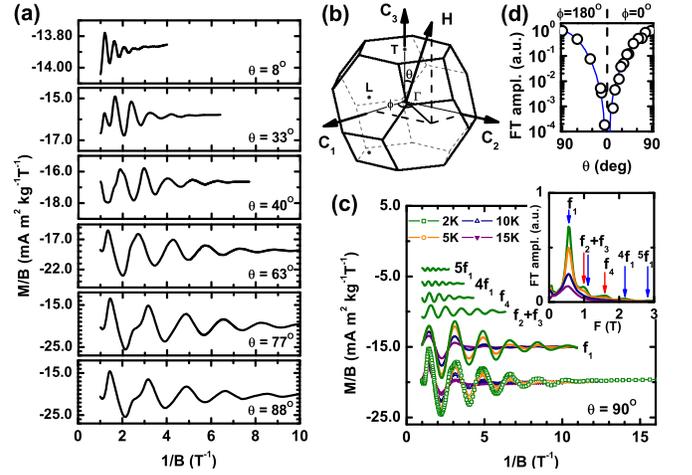}
\caption{(Color online) 
(a) Example of the angular dependence of the dHvA oscillations in
Bi$_{0.91}$Sb$_{0.09}$ at 2 K for magnetic
fields in the binary plane ({\it i.e.}, $\theta$ is varied, while
$\phi$ = 0$^{\circ}$ is fixed). 
(b) Schematic picture of the bulk
Brillouin zone of Bi$_{1-x}$Sb$_x$ and the geometry of the experiment.
(c) Example of the decomposition of the raw $M/B$ data (shown by symbols
at the bottom) into a set of oscillatory parts (shown by straight
lines); inset shows the FT spectra at 2, 5, 10, and 15 K for $\theta$ = 
90$^{\circ}$ and $\phi$ = 180$^{\circ}$. 
(d) $\theta$ dependence of the FT amplitude of the main peak $f_1$ 
within the binary plane. 
}
\label{fig2}
\end{figure}

The dHvA effect, which we also observe in our Bi$_{0.91}$Sb$_{0.09}$
crystal, is another way of probing the Fermi surface. This effect is
manifested in magnetization which is a very straightforward
thermodynamic property. An example of oscillations of the magnetic
susceptibility $M/B$ measured at 2 K with the magnetic field direction
lying in the binary plane (perpendicular to the $C_2$ axis) at different
angles $\theta$ is shown in Fig. 2(a), where a complex structure of the
oscillations as well as changes in their amplitudes and periods are
clearly seen. Shown in Fig. 2(b) is a schematic representation of the
BZ, three main axes, positions of high symmetry points, and geometry of
the present experiment. The Fourier analysis is applied to all measured
data to obtain quantitative information, and Fig. 2(c) shows, as an
example, the decomposition of magnetic oscillations into a set of
constituents for $B \parallel C_1$ axis. The experimental data are shown
at the bottom by symbols. The inset of Fig. 2(c) displays the FT spectra
for four measured temperatures; several frequencies can be readily
distinguished, and up to the fifth harmonic of the lowest frequency
$f_1$ can be identified at 2 K, indicating a high degree of coherence
\cite{Fal}. The angular dependence of the FT amplitude of the main peak
$f_1$ for magnetic fields in the binary plane is shown in Fig. 2(d).
This amplitude exhibits a very sharp decrease by as much as four orders
of magnitude when the magnetic-field direction approaches the bisectrix
plane (perpendicular to the $C_1$ axis), as if carriers cannot move
perpendicular to this plane to make Larmor orbit. This angular
dependence is the first indication of a two-dimensional (2D) nature of
the corresponding FS.

\begin{figure}\includegraphics*[width=21pc]{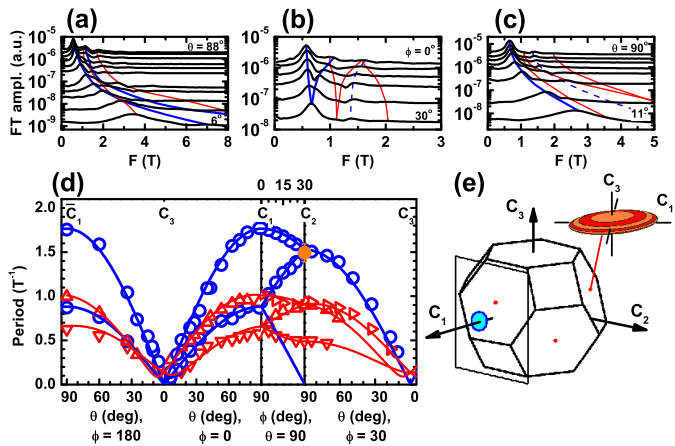}
\caption{(Color online) 
(a)--(c) Evolution of the FT components of the dHvA oscillations
upon rotating the magnetic field within the (a) binary,
(b) trigonal, and (c) bisectrix planes. The lines are guides to the eye, 
and the dashed lines denote those peaks that are not independent but 
come from higher harmonics or interference of independent frequencies. 
(d) Angular dependences of $P$ (= $1/F$) within the three main planes; 
experimental data,
corresponding to the peaks in (a)--(c), are shown by symbols. Thick
(blue) and thin (red) solid lines are the fits using the
model considering the 2D and 3D Fermi surfaces, respectively (see text).
(e) Schematic picture of the surface and bulk FSs and their
positions in the BZ.
}
\label{fig3}
\end{figure}

Since the frequencies of QOs, $F$, are directly related to the
Fermi-surface cross sections via the Onsager relation $F = (\hbar c / 2
\pi e)A$, where $A$ is the area of an extremal cyclotron orbit, the
shape and size of the FS can be obtained from the angular dependences of
$F$ measured within the three main crystallographic planes. To separate
the FT peaks associated with different Fermi surfaces, the
angle-dependent FT spectra in the binary, trigonal, and bisectrix planes
(which are perpendicular to the $C_2$, $C_3$, and $C_1$ axes,
respectively) has been mapped [Figs. 3(a)--3(c)]. By following the
change in position of individual peaks in the spectra with changing
angle, as shown in Figs. 3(a)--3(c) by lines, it is easy to sort out the
evolution of each peak, which allows us to plot the angular dependences
of all periods $P = 1/F$ of the dHvA oscillations within the three main
planes, as shown in Fig. 3(d) by symbols. The period of the SdH
oscillations [Fig. 1(c)] is also shown by a filled circle in Fig. 3(d).
The branches in Fig. 3(d) are separated into two sets: In the binary
plane ($\bar{C_1}-C_3-C_1$), two branches (shown by open circles) are
symmetrical with respect to the positive ($C_1$) and negative
($\bar{C_1}$) axis direction and have a minimum (in fact, approaching
zero) at exactly $\theta$ = 0$^{\circ}$, while the other two branches
(shown by open triangles) are slightly shifted with respect to $\theta$
= 0$^{\circ}$. 

Taking into account all three planes, the angular dependences of the
observed periods point to a coexistence of 2D and 3D Fermi surfaces: The
2D one is a circle of radius $k_{F}$ = 4.15$\times$10$^{7}$ m$^{-1}$ in
the surface BZ of the bisectrix (2$\bar{1} \bar{1}$) surface, which is
perpendicular to the $C_1$ axis, and there are six of them by symmetry.
The size of this circle corresponds to the surface charge-carrier
concentration $n_s$ = $k_{F}^{2}/4 \pi$ = 1.4$\times$10$^{14}$ m$^{-2}$,
if the surface state is spin-filtered; it will be twice as large, if the
state is not spin-filtered. Branches calculated in this model are shown
in Fig. 3(d) by thick (blue) solid lines, which match the data almost
completely. The 3D FS is a set of three ellipsoids, located at the $L$
points of the BZ with semi-axes $a$ = 2.7$\times$10$^{8}$ m$^{-1}$, $b$
= 1.3$\times$10$^{8}$ m$^{-1}$, and $c$ = 2.3$\times$10$^{7}$ m$^{-1}$
that are approximately along $C_1$, $C_2$, and $C_3$, respectively; to
be precise, the ellipsoids are tilted by $\sim$+6$^{\circ}$ in the
binary planes [see Fig. 3(e)]. The size of each ellipsoid corresponds to
the bulk charge carrier concentration of $abc/3 \pi^{2}$ =
2.7$\times$10$^{22}$ m$^{-3}$. In total, three ellipsoids give
$\sim$8.1$\times$10$^{22}$ m$^{-3}$. Calculated branches for those 3D
states are shown in Fig. 3(d) by thin (red) solid lines, which, again,
match the data very well. The shapes of the 2D and 3D FSs are
schematically shown in Fig. 3(e). 

An important parameter that can be extracted from the dHvA oscillations
is the cyclotron mass $m_c$. The temperature dependences of the
oscillation amplitude for different angle $\theta$ (with fixed $\phi$ =
0$^{\circ}$) are shown in Fig. 4(a), where fittings of the data to the
standard Lifshitz-Kosevich theory \cite{Schnbrg} are shown by solid
lines \cite{note1}. This analysis gives the angular dependence of $m_c$
associated with the surface states as shown in Fig. 4(b). Two prominent
features are readily recognized: First, the absolute value of $m_c$ in
the magnetic field perpendicular to the surface ($\theta$ =
90$^{\circ}$) is extremely small, measuring only 0.0057$m_e$ ($m_e$ is
the free electron mass), which is smaller than the cyclotron mass in
pure Bi for any direction. Such a small $m_c$ is the reason why it is
possible to observe pronounced oscillations at relatively low magnetic
fields. Second, $m_{c}(\theta)$ diverges at $\theta$ = 0$^{\circ}$ and
shows the $1/\sin\theta$ dependence that is characteristic of a 2D FS,
giving unambiguous evidence for the 2D nature \cite{note2}.

\begin{figure}\includegraphics*[width=21pc]{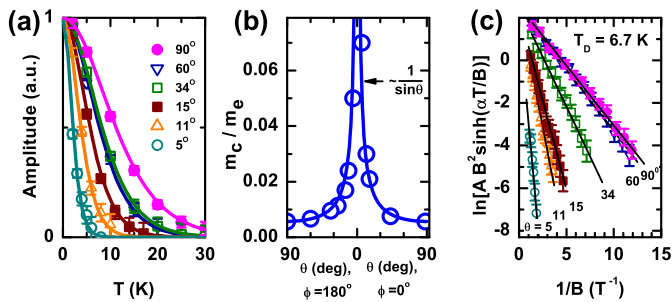}
\caption{(Color online) 
(a) Temperature dependences of the dHvA amplitudes measured at a fixed
magnetic field below the quantum limit within the binary plane. Solid
lines are the fits to the standard Lifshitz-Kosevich formula. 
(b) Angular dependence of $m_c$ signifying the unambiguous 2D nature.
(c) Dingle plots yielding the same Dingle temperature for all 
magnetic-field directions ($A$ is the dHvA amplitude and $\alpha$ = 
$14.7(m_c/m_e)$ [T/K]).
}
\label{fig4}
\end{figure}

Once $m_c$ is known, the scattering time $\tau$ is obtained by analyzing
how $M/B$ changes with $B^{-1}$ at fixed temperatures. Figure 4(c) shows
the Dingle plots \cite{Schnbrg}, which yields the Dingle temperature
$T_D$ (= $\hbar / 2 \pi \tau k_{B}$), for the same field directions as
in Fig. 4(a). All curves show linear slopes and for all angles the same
$T_D$ of 6.7 K (which corresponds to $\tau \approx$
1.8$\times$10$^{-13}$ s) is obtained. This is surprisingly low effective
temperature for an alloy with this doping concentration \cite{Schnbrg}.
Since the Fermi velocity $v_{F}$ can be approximated by $v_{F}$ = $\hbar
k_{F} / m_{c} \approx$ 8.5$\times$10$^{5}$ m/s, the mean free path
$\ell$ on the surface is estimated as $\ell$ = $v_{F} \tau \approx$ 150
nm, which is comparable to the de Broglie wave length $\lambda$ = 2$\pi
/ k_{F} \approx$ 150 nm for the 2D FS. In passing, we note that our
samples have a rectangular shape and four of the surfaces were cut along
the $C_3$ axis with a wire saw; hence, those four surfaces are actually
a collection of locally-flat mini-surfaces with various orientations.
Naturally, some of the mini-surfaces are perpendicular to the $C_1$
axis, and as long as their extent is larger than $\ell$, they can produce
QOs.

Let us now combine the dHvA results on Bi$_{0.91}$Sb$_{0.09}$ with the
transport data. The bulk FS gives the total number of charge carriers,
$n_{\rm bulk}$, of $\sim$8.1$\times$10$^{22}$ m$^{-3}$, which is close
to the value $\sim$1.8$\times$10$^{23}$ m$^{-3}$ estimated from $R_H$.
This means that the observed bulk ellipsoids must be electron pockets.
The band scheme \cite{Len} of Bi$_{1-x}$Sb$_{x}$ dictates that those
electron pockets be located at the three $L$ points in the BZ, which is
consistent with our model to fit the data in Fig. 3(d). The surface FS
is most likely a hole pocket, because the measured $R_H$ gives an
overestimated carrier concentration which is naturally understood if
there is a partial cancelation of the Hall effect from the electron and
hole contributions. 

We note that $n_{\rm bulk}$ obtained from the dHvA effect is much larger
than the possible concentration of impurities, donors or acceptors, of
$\sim$10$^{21}$ m$^{-3}$ (estimated from the ICP-AES analysis) in our
Bi$_{0.91}$Sb$_{0.09}$ crystal. Then, why is the Fermi level in the bulk
of Bi$_{0.91}$Sb$_{0.09}$ not located in the gap but goes into the
conduction band, giving a rather appreciable electron concentration? We
speculate that this is due to a coupling of the bulk FS to the surface
FS that causes an intrinsic self-doping, but the actual mechanism is to
be elucidated. Also, the amplitude of the dHvA oscillations from the 2D
FS appears to be orders of magnitude larger than what is expected if
each surface hole contributes a magnetic moment of $\sim \hbar e/m_c c$;
this puzzle will probably be resolved if the unique properties of the
surface states of a TI, the Dirac-like \cite{ShGu} and spin-filtered
nature and the topological protection, are properly taken into account
\cite{note3}. An alternative scenario that may account for the large
dHvA amplitudes from the 2D FS and the unusual electron-hole coupling
would be a formation of a 2D cylindrical FS perpendicular to the
bisectrix planes within the bulk BZ; however, this is probably too
exotic, given the 3D crystal structure of Bi$_{1-x}$Sb$_x$. Clearly,
there is a lot to understand about the macroscopic properties of
topological insulators, and the availability of the powerful tool of
quantum oscillations demonstrated here promises a vast possibility for
future research. 

\begin{acknowledgments}
We thank M. Hagiwara, Y. Onuki, and S.-C. Zhang for helpful discussions. 
This work was supported by JSPS (KAKENHI 19340078 and 2003004) and AFOSR 
(AOARD-08-4099).
\end{acknowledgments}


\begin{thebibliography}{99}
\bibitem{note1}
Maximums, that are related to the crossing of the Fermi energy by
Landau levels with higher occupation numbers, were chosen for this
analysis to reduce contributions from frequencies other than $f_1$.
\bibitem{note2}
The angular dependence $m_{c}(\phi)$
in the trigonal plane is also close to $\sim 1/\sin(90^{\circ}-\phi)$,
but the span of $\phi$ is limited to 0$^{\circ}$--30$^{\circ}$ due to
the crystal symmetry, precluding a divergence in $m_{c}(\phi)$.
\bibitem{note3} 
In magnetic fields, the Kramers degeneracy is lifted and a Zeeman gap
opens in the topological surface state. But the surface can remain
metallic as long as the Fermi level is located within the surface band.
\end{thebibliography}
\end{document}